\documentclass{PoS}
\PoS{PoS(WC2004)006}
\title{11D Supergravity as a Gauge Theory for the M-Algebra}
\ShortTitle{11D Supergravity as a Gauge Theory for the M-Algebra}
\author{Mokhtar Hassa\"{\i}ne,\\
        CECS, Valdivia, Chile\\
        E-mail: \email{hassaine-at-cecs.cl}}
\author{Ricardo Troncoso\\
        CECS, Valdivia, Chile\\
        E-mail: \email{ratron-at-cecs.cl}}
\author{\speaker{Jorge Zanelli}\\
        CECS, Valdivia, Chile\\
        E-mail: \email{jz-at-cecs.cl}}

\abstract{The eleven-dimensional gravitational action invariant under local
Poincar\'{e} transformations is given by the dimensional continuation of the
Euler class of ten dimensions. Here we show that the supersymmetric extension
of this action leads, through the Noether procedure, to a theory with the local
symmetry group given by the \emph{M}-algebra. The fields of the theory are the
vielbein $e^a_{\mu}$, the Lorentz (spin) connection $\omega^{ab}_{\;\mu}$, one
gravitino ($\psi_{\mu}$), and two 1-forms, $b^{ab}_{\mu}$ and
$b^{abcde}_{\mu}$, which transform as antisymmetric Lorentz tensors. These
fields are components of a single connection for the \emph{M}-algebra and the
supersymmetric Lagrangian can be seen to be a Chern-Simons form. The dynamics
has a multiplicity of degenerate vacua without propagating degrees of freedom.
The theory is shown to admit solutions of the form $S^{10-d}\times X_{d+1}$,
where $X_{d+1}$ is a warped product of $\Bbb{R}$ with a $d$-dimensional
spacetime. Among this class, the gravitational effective action describes a
propagating graviton only if $d=4$ and the spacetime has positive cosmological
constant. The perturbations around this solution reproduce linearized General
Relativity around four-dimensional de Sitter spacetime.}

\FullConference{Fourth International Winter Conference on Mathematical Methods
in Physics\\
09 - 13 August 2004\\
Centro Brasileiro de Pesquisas Fisicas (CBPF/MCT), Rio de Janeiro, Brazil}

\begin{document}

\section{Introduction} 

Ever since Theodor Kaluza \cite{Kaluza} and Oskar Klein \cite{Klein}
conjectured the possibility that four-dimensional spacetime could be a section
of a larger manifold where all laws of physics would be unified, theoretical
physicists have tried to realize this dream with varying degrees of success.
This is such a compelling idea, that many of us believe it must be true at some
level, but we have mostly failed to pinpoint exactly how it may work. Among the
most trodden paths one finds the attempt to describe the higher--dimensional
spacetimes as Einstein did with our own: assuming it faithfully described by
the Einstein-Hilbert (\textbf{EH}) action. It turns out that under reasonable
assumptions, the Einstein theory is unique --modulo the cosmological constant--
only in four dimensions, but is one very particular option among many in higher
dimensions. In this lecture, we analyze some consequences of freeing ourselves
from the straitjacket of the Einstein-Hilbert action.  The result in this case
is not, as one may fear, opening a Pandora's box of unlimited possibilities,
but a controlled and very restricted set of theories known as Lovelock
gravities \cite{Lovelock}. These theories extend Einstein's General Relativity,
and a very narrow subset of them is known to accept simple supersymmetric
extensions\footnote{These are the Chern-Simons (\textbf{CS}) theories of
gravity, which exist in dimensions $D=2n+1$, and possess local off-shell
invariance under the $D$-dimensional anti-de Sitter group $SO(D-1,2)$, or its
contraction, the Poincar\'{e} group, $ISO(D-1,1)$.}.

The standard procedure to carry out the Kaluza-Klein dream is to start with a
theory in a sufficiently high dimension whose gravitational sector is given by
the EH action. Then, the four-dimensional physical world is expected to arise
from compactification of the extra dimensions (see, e.g.,
\cite{Duff-Nilsson-Pope}), or through some more recent alternatives
\cite{Alternative}. This is extremely reasonable and straightforward. The
problem is that the fact that the low-energy behavior of the universe is
four--dimensional is an assumption rather than a prediction of the theory. This
is hardly convincing, specially since the gravitational part of the action
should be in charge of determining the properties of the spacetime geometry
including, presumably, the explanation for its effective four--dimensional
appearance.

A manifestation of this problem is the following paradox: since eleven
dimensional Minkowski space is a maximally (super)symmetric state, and the
theory is well behaved around it, then, what prompts the space to curl up
choosing a particular compactification with less symmetry as the vacuum? And,
why not six instead of four dimensions? Ideally, the eleven-dimensional theory
should dynamically predict a low energy regime corresponding to a
four-dimensional effective theory. In such scenario, a background solution with
an effective spacetime dimension greater than four should be expected to be
some sort of false vacuum.

The origin of the problem lies in the fact that the dynamics produced by the EH
action is insensitive to dimensionality in the sense that its linearized
approximation yields a well behaved wave equation for all dimensions (greater
than three). Clearly, the solutions of the equations depend on $D$ --in
particular, black holes, planetary orbits and life itself would be very
different if $D\neq4$--, but there is no compelling reason, within the theory,
why $D$ \emph{should be} $4$.

Here we discuss a theory defined in eleven--dimensional spacetime whose
gravitational sector is singled out by requiring the theory to be locally
invariant under the Poincar\'{e} group.  With this assumption one can go a long
way to specify the entire locally supersymmetric extension and to study its
physical consequences. In particular, after supersymmetrization the theory
turns out to be invariant under the symmetry group generated by the the maximal
extension of the ${\cal N}=1$ super-Poincar\'{e} algebra in eleven dimensions,
commonly known as \emph{M}-algebra. This algebra is spanned by the set of
generators $G_A=\{ J_{ab}, P_{a}, Q_{\alpha}, Z_{ab}, Z_{abcde} \}$, where
$J_{ab}$ and $P_{a}$ are the generators of the Poincar\'{e} group and
$Q_{\alpha }$ is a Majorana spinor supercharge with
anticommutator\footnote{Note that, contrary to the case in standard
supergravity, the generators of diffeomorphisms ($\mathcal{H}_{\mu }$) are
absent from the right hand side of (\ref{MAlgebra}).} \cite{Townsend-L}
\begin{equation}
\{Q_{\alpha },Q_{\beta }\}=\left( C\Gamma ^{a}\right) _{\alpha \beta
}P_{a}+(C\Gamma ^{ab})_{\alpha \beta }Z_{ab}+(C\Gamma ^{abcde})_{\alpha \beta
}Z_{abcde}\;.  \label{MAlgebra}
\end{equation}

The charge conjugation matrix $C$ is antisymmetric, and the generators $Z_{ab}$
and $Z_{abcde}$ are tensors under Lorentz rotations, but otherwise Abelian.
This algebra is expected to generate the symmetries of an underlying
fundamental theory in eleven dimensions known as M-Theory
\cite{Townsend-D,Witten,Schwarz}.

Our choice of eleven dimensions may be justified by the wish to explore
geometrical and dynamical structures that could be regarded as new ``cusps'' of
the M-theory diagram (see e.g., \cite{WNS}), in particular because the theory
presented here is not equivalent to the standard Cremmer-Julia-Scherk
supergravity in eleven dimensions \cite{CJS}. Other dimensions and other
symmetry groups --such as the supersymmetric extensions of the AdS group-- are
also reasonable alternatives.

As shown below, the locally supersymmetric extension of the Poincar\'{e}
invariant action fixes the field content to include, apart from the graviton
$e_{\mu }^{a}$ , the spin connection $\omega _{\mu }^{ab}$ and the gravitino
$\psi _{\mu }$, two one-form bosonic fields $b_{\mu }^{ab}$, $b_{\mu}^{abcde}$,
which are rank two and five antisymmetric tensors under the Lorentz group,
respectively. The local supersymmetry algebra closes off-shell, without need
for auxiliary fields.

As will be shown, the supersymmetric Lagrangian can be explicitly written as a
Chern-Simons form. It is known that for Chern-Simons theories bosonic and
fermionic degrees of freedom do not necessarily match, since the dynamical
fields are assumed to belong to a connection instead of a multiplet for the
supergroup \cite{Tr-Z} and, as it occurs in three dimensions, there exists an
alternative to the introduction of auxiliary fields (see e. g. \cite{HIPT}).

\section{Gravitational sector} 

General Relativity in dimensions higher than four is generalized by the
so-called Lovelock theories of gravity. These theories respect the assumptions
of general covariance, second order field equations for the metric, and they
include the Einstein-Hilbert lagrangian as a particular case \cite {Lovelock}.
The Lovelock Lagrangians are linear combinations of the dimensional
continuations of the Euler densities from all lower dimensions \cite{Zumino},
namely,
\begin{equation}\label{Lovelock}
    L=\sum_{p=0}^{\left[\frac{D-1}{2}\right]}\alpha_p L_p,
\end{equation}
with
\begin{equation}\label{Lp}
    L_p= \epsilon_{a_1 \cdots a_{D}} R^{a_1 a_2}\cdots R^{a_{1p-1}
    a_{2p}}\wedge e^{a_{2p+1}} \cdots e^{a_D},
\end{equation}
where $R^{a}_{\;b}=d\omega^{a}_{\;b}+ \omega^{a}_{\;c}\omega^{c}_{\;b}$ is the
curvature two-form (wedge products between forms is understood throughout), and
$\alpha_p$ are arbitrary coefficients. In the series (\ref{Lovelock}), $L_0$
corresponds to the cosmological constant, $L_1$ gives the Einstein-Hilbert
Lagrangian and $L_2$ is the Gauss-Bonnet term. The Lovelock Lagrangians are by
construction invariant under diffeomorphisms and local Lorentz transformations,
but in odd dimensions there are special choices of the coefficients $\alpha_p$
for which the Lovelock Lagrangians acquire a larger local symmetry:
Poincar\'{e} or (anti-)de Sitter, as in the next table.

\begin{center}
\begin{tabular}{|l|l|}
  \hline
  $\alpha_p$ & Local symmetry ($D=2n+1$)\\
  \hline
 Arbitrary & Lorentz $\;\;\;\;\;\;\;\;\;\;\;  SO(D-1,1)$, [also for even $D$] \\
 $\delta^n_{\;p}$ & Poincar\'{e} $\;\;\;\;\;\;\;\;\;  ISO(D-1,1)$\\
  $\frac{l^{2p-D}}{D-2p}$ $\left(\begin{array}{c}
    n \\ p\end{array}\right)$ & anti-de Sitter $\;\;SO(D-1,2)$ \\
  $(-1)^p\frac{l^{2p-D}}{D-2p}$ $\left(\begin{array}{c}
    n \\ p\end{array}\right)$  &de Sitter $\;\;\;\;\;\;\;\;\;\; SO(D,1)$ \\
  \hline
\end{tabular}
\end{center}

Since the action can be expressed in terms of differential forms without using
the Hodge dual, these theories cannot yield higher order field equations: In
the first order formalism (analogous to Palatini's) the field equations can
only involve the first exterior derivatives of the dynamical fields (see, e.g.,
\cite{Troncoso-Zanelli}). Then, if the vanishing torsion condition is imposed,
the field equations become at most second order. In the vanishing torsion
sector, the theory has the same degrees of freedom as General Relativity
\cite{Teitelboim-Zanelli}. If the torsion constraint is not imposed, the field
equations remain first order, even if the theory were coupled to other $p$-form
fields without involving the Hodge. In fact, in this way it is impossible to
generate higher derivative terms in this theory.

An action with local Poincar\'{e} symmetry must be, in particular, invariant
under local translations. Since the only field with the right tensor structure
to transform as a connection under translations is the vielbein, an
infinitesimal local translation must act on the fields as
\begin{equation}
\delta e^{a}=D\lambda^{a}=d\lambda^{a}+\omega_{\,b}^{a}\lambda^{b},\;\; \;\;\;
\delta \omega^{ab}=0\;.  \label{PoincareTrans}
\end{equation}
The unique Lovelock lagrangian invariant under local translations is
\cite{Troncoso-Zanelli,Chamseddine},
\begin{equation}
I_{G}[e,\omega ]=\int_{M_{11}}\epsilon _{a_{1}\cdots
a_{11}}R^{a_{1}a_{2}}\cdots R^{a_{9}a_{10}}e^{a_{11}},  \label{I_G}
\end{equation}
which corresponds to choosing the coefficients $\alpha_p=\delta^n_{\;p}$.

For the reason given above, we take $I_{G}$ as the gravitational sector of our
theory rather than the Einstein-Hilbert action, which is not invariant under
(\ref{PoincareTrans}) \footnote{Under the transformation (\ref{PoincareTrans}),
the Einstein-Hilbert action $I_{EH}=\int \epsilon_{a_1 \cdots a_{11}} R^{a_1
a_2} e^{a_3} \cdots e^{a_{11}}$ changes by a term proportional to $\int
\epsilon_{a_1 \cdots a_{11}} R^{a_1 a_2} T^{a_3}e^{a_4} \cdots e^{a_{10}}
\lambda^{a_{11}}$, which vanishes only if the torsion equation $T^a = De^a=0$
is used. However, this last condition is incompatible with the transformations
because $\omega^{ab}$ would be a function of $e^a$ and hence, $\delta
\omega^{ab}$ would not vanish. See \cite{Regge}}. The Lagrangian in (\ref{I_G})
is the ten--dimensional Euler density continued to eleven dimensions and
contains the degrees of freedom of eleven dimensional gravity
\cite{Teitelboim-Zanelli}.

A local Poincar\'{e} transformation acting on the dynamical fields is a gauge
transformation $\delta _{\lambda }A=d\lambda +[A,\lambda ]$, with parameter
$\lambda =$ $\lambda ^{a}P_{a}+\frac{1}{2}\lambda ^{ab}J_{ab}$, provided
$e^{a}$ and $\omega ^{ab}$ are the components of a single connection for the
Poincar\'{e} group, $A=e^{a}P_{a}+\frac{1}{2}\omega ^{ab}J_{ab}.$ This
observation will be the guiding principle for the construction of a locally
supersymmetric extension of $I_{G}$.

\section{Supersymmetric extension}  

A natural way to construct the locally supersymmetric extension of (\ref{I_G})
without breaking local Poincar\'{e} invariance is by demanding that the new
fields required by supersymmetry enter on the same footing as the original
ones. In other words, all dynamical fields will be assumed to belong to a
single connection for a supersymmetric extension of the Poincar\'{e} algebra.
The simplest option would be to consider the $\mathcal{N}=1$ super Poincar\'{e}
algebra without central extensions. However, this\ possibility must be ruled
out. Indeed, in this case, the connection would be extended by the addition of
a gravitino as $A\rightarrow A+\psi Q$, and the gauge generator would change as
$\lambda \rightarrow \lambda +\epsilon Q$, where $\epsilon $\ is a zero-form
Majorana spinor. This fixes the supersymmetric transformations to be
\begin{equation}\label{SS1}
\delta_{\varepsilon}e^{a}=\bar{\epsilon}\Gamma^{a}\psi, \;\;\;
\delta_{\varepsilon}\psi=D\epsilon, \;\;\; \delta_{\varepsilon}\omega^{ab}=0.
\end{equation}
Then, the variation of (\ref{I_G}) under supersymmetry is canceled by a kinetic
term for the gravitino of the form
\begin{equation}
I_{\psi }=-\frac{1}{3}\int_{M_{11}}R_{abc}\bar{\psi}\Gamma ^{abc}D\psi ,
\label{I-Psi}
\end{equation}
where $R_{abc}:=\epsilon _{abca_{1}\cdots a_{8}}R^{a_{1}a_{2}}\cdots
R^{a_{7}a_{8}}$. The variation of $I_{\psi }$ produces, in turn, an extra piece
which cannot be canceled by a local Lagrangian for $e^{a}$, $\omega ^{ab}$, and
$\psi $, and hence the super Poincar\'{e} algebra is not rich enough to ensure
the off-shell supersymmetry of the action.

On the other hand, following the Noether procedure, it can be seen that
supersymmetry may be achieved if additional bosonic fields are introduced.
These fields can only be a second-rank or a fifth-rank tensor one-forms
$b^{ab}$, and $b^{abcde}$, which transform as
\begin{equation}
\begin{array}{lr}
\delta_{\varepsilon}b^{ab}= \bar{\epsilon}\Gamma^{ab}\psi , &\;\;
\delta_{\varepsilon}b^{abcde}= \bar{\epsilon}\Gamma^{abcde}\psi ,
\end{array}
\label{SS2}
\end{equation}
respectively.

Assuming that the dynamical fields belong to a single connection for a
supersymmetric extension of the Poincar\'{e} group, the only option that brings
in these extra bosonic fields is to consider the M-algebra (\ref{MAlgebra}).
Additionally, this also prescribes their supersymmetry transformations in the
expected form (\ref{SS2}). This means that the field content is given by the
components of a single fundamental field, the M-algebra connection,
\begin{equation}
A=\frac{1}{2}\omega ^{ab}J_{ab}+e^{a}P_{a}+ \psi ^{\alpha }Q_{\alpha
}+b^{ab}Z_{ab}+b^{abcde}Z_{abcde}\;,  \label{M-connection}
\end{equation}
and the required local supersymmetry transformations (\ref{SS1}, \ref{SS2}) are
obtained from a gauge transformation of the M-connection (\ref{M-connection})
with parameter $\lambda =\epsilon^{\alpha}Q_{\alpha}$.

Thus, the supersymmetric extension of (\ref{I_G}) is found to be
\begin{eqnarray}
I_{\alpha } &=&I_{G}+I_{\psi}-\frac{\alpha}{6}\int_{M_{11}}R_{abc}R_{de}
b^{abcde}  \nonumber \\ &&+16(1-\alpha )\int_{M_{11}}[R^{2}R_{ab}-6(R^3)_{ab}]
R_{cd} \left( \bar{\psi}\Gamma^{abcd} D\psi -6R^{[ab}b^{cd]}\right),
\label{Action}
\end{eqnarray}
where $R^{2}:=R^{ab}R_{ba}$ and $(R^{3})^{ab}:=R^{ac}R_{cd}R^{db}$. Here $%
\alpha $ is a dimensionless constant whose meaning will be discussed below.

This action is invariant under (\ref{PoincareTrans}, \ref{SS1}, \ref{SS2}),
local Lorentz rotations, and also under the local Abelian transformations
\begin{equation}
\begin{array}{ll}
\delta b^{ab}=D\theta ^{ab}, & \delta b^{abcde}=D\theta ^{abcde}
\end{array}
.  \label{ZZ}
\end{equation}
Invariance under general coordinate transformations is guaranteed by the use of
exterior forms. It is simple to see that the local invariances of the action,
including Poincar\'{e} transformations, supersymmetry and the Abelian
transformations (\ref{ZZ}), are obtained by a gauge transformation of the
M-connection (\ref{M-connection}) with parameter $\lambda =\lambda^{a} P_{a} +
\frac{1}{2}\lambda^{ab}J_{ab}+\theta^{ab}Z_{ab}+\theta^{abcde}Z_{abcde}+
\epsilon^{\alpha} Q_{\alpha}$. As a consequence, the off-shell invariance of
the action under the supersymmetry algebra is ensured by construction without
invoking field equations or requiring auxiliary fields.

\subsection{Manifest M-covariance} 

The action (\ref{Action}) describes a gauge theory for the M-algebra with fiber
bundle structure, which can be seen explicitly by writing the Lagrangian as a
Chern-Simons form \cite{FootnoteCS} for the M-connection (\ref{M-connection}).
Indeed, the Lagrangian satisfies $dL=\left\langle F^{6}\right\rangle $, where
the curvature $F=dA+A^{2}$ is given by
\[
F=\frac{1}{2}R^{ab}J_{ab}+\tilde{T}^{a}P_{a}+ D\psi ^{\alpha }Q_{\alpha
}+\tilde{F}^{^{[2]}}Z_{^{[2]}}+\tilde{F}^{^{[5]}}Z_{^{[5]}},
\]
with $\tilde{T}^{a}=De^{a}-(1/2)\bar{\psi}\Gamma^{a}\psi $ and $\tilde{F}^{[k]}
=Db^{[k]}-(1/2)\bar{\psi}\Gamma ^{[k]}\psi $ for$\,\,k=2$ and $5$. The bracket
$\left\langle ...\right\rangle $ stands for a multilinear form of the M-algebra
generators $G_{A}$ whose explicit expression is far from obvious. In the case
at hand, it can be shown that the only nonvanishing components of the bracket
are given by
\[
\begin{array}{l}
\left\langle J_{a_{1}a_{2}},\cdots ,J_{a_{9}a_{10}},P_{a_{11}}\right\rangle =
\frac{16}{3}\epsilon _{a_{1}\cdots a_{11}}\;, \\
\left\langle J_{a_{1}a_{2}},\cdots ,J_{a_{9}a_{10}},Z_{abcde}\right\rangle
=-\alpha \frac{4}{9}\epsilon _{a_{1}\cdots a_{8}abc}\eta
_{[a_{9}a_{10}][de]}\;, \\
\left\langle
J_{a_{1}a_{2}},J_{a_{3}a_{4}},J_{a_{5}a_{6}},J^{a_{7}a_{8}},J^{a_{9}a_{10}},
Z^{ab}\right\rangle =(1-\alpha )\frac{16}{3}\left[ \delta _{a_{1}\cdots \cdots
a_{6}}^{a_{7}\cdots a_{10}ab}-\delta _{a_{1}\cdots a_{4}}^{a_{9}a_{10}ab}\delta
_{a_{5}a_{6}}^{a_{7}a_{8}}\right]  \\
\left\langle
Q,J_{a_{1}a_{2}},J^{a_{3}a_{4}},J^{a_{5}a_{6}},J^{a_{7}a_{8}},Q\right\rangle
=\frac{32}{15}\left[ C\Gamma _{a_{1}a_{2}}^{\,\,\quad
\;\,a_{3}\cdots a_{8}}+\right.  \\
\;\;\;\;\;\;\;\;\;\;\;\;\;\;\;\;\;\;\;\;\;\;\;\;\;\;\;\;\;\;\;\;\;\;\;\;\;\;%
\;\;\;\qquad \;\;\;\;\;\;\;\;\left. (1-\alpha )\left( 3\delta
_{a_{1}a_{2}ab}^{a_{3}\cdots a_{6}}C\Gamma ^{a_{7}a_{8}ab}+2C\Gamma
^{a_{3}\cdots a_{6}}\delta _{a_{1}a_{2}}^{a_{7}a_{8}}\right) \right] ,
\end{array}
\]
where (anti-)symmetrization under permutations of each pair of generators is
understood when all the indices are lowered\footnote{In general, without an
explicit representation of the algebra, it is a highly nontrivial exercise to
determine the existence and the form of an invariant tensor of arbitrary
rank.}.  The explicit form of this bracket allows writing the field equations
in a manifestly covariant form as
\begin{equation}
\left\langle F^{5}G_{A}\right\rangle =0.  \label{FieldEqs}
\end{equation}
In addition, if the eleven-dimensional spacetime is the boundary of a
twelve-dimensional manifold, $\partial \Omega _{12}=M_{11}$, the action (\ref
{Action}) can also be\ written as $I=\int_{\Omega _{12}}\left\langle
F^{6}\right\rangle $, which describes a topological theory in twelve
dimensions. In spite of its topological origin, the eleven--dimensional action
does possess propagating degrees of freedom and hence it should not be thought
of as a topological field theory.

\section{Gravitons and four-dimensional spacetime} 
We now turn to the problem of identifying one true vacuum of the theory.
Obviously, a configuration with a locally flat connection, $F=0$, solves the
field equations (\ref{FieldEqs}) and would be a natural candidate for vacuum in
a standard field theory. Moreover, such state is invariant under all gauge
transformations and is, therefore, maximally supersymmetric under (\ref{SS1})
and (\ref{SS2}) which makes it likely to be a stable (BPS) configuration.

The identification of this solution with a vacuum state is more compelling in
view of the fact that it carries no charges of any kind and invariant under
spacetime transformations and supersymmetry. Matter-free eleven-dimensional
Minkowski spacetime is an example of this. However, no local degrees of freedom
can propagate on such background: all perturbations around it are zero modes.

In a matter-free configuration, $\psi=0$, $b^{[2]}=0$, $b^{[5]}=0$, Eq.
(\ref{FieldEqs}) is a set of polynomial equations of fifth degree in the
curvature two-forms. In particular, the equations obtained varying with respect
to the vielbein and the spin connection are
\begin{equation}\label{R5}
    \epsilon_{A A_1\cdots A_{10}}R^{A_1A_2} \cdots R^{A_9A_{10}}=0
\end{equation}
\begin{equation}\label{R4T}
    \epsilon_{A B A_1\cdots A_{9}}R^{A_1A_2} \cdots R^{A_7A_8} T^{A_9}=0.
\end{equation}
Thus, in order to have a propagating connection, the spatial components
$R^{AB}_{\;\;ij}$ cannot be small. Alternatively, a deviation that propagates
around a flat background $R^{AB}=0$ cannot be infinitesimal and is therefore
non-perturbative.  Moreover, since the derivatives of the field cannot be small
either, the deviations are necessarily non-local. This feature is not altered
by the remaining equations. Alternatively, in order to have well-defined
linearized perturbations, a background solution must be a simple zero of one of
the set of equations. In particular, this requires the curvature to be
nonvanishing on a submanifold of a large enough dimension.

\subsection{Nontrivial vacuum geometry} 

Let us consider a torsionless spacetime with a product geometry of the form
$X_{d+1}\times S^{10-d}$, where $X_{d+1}$ is a domain wall whose worldsheet is
a $d$-dimensional spacetime $M_{d}$. The line element is given by
\begin{equation}
ds^{2}=e^{-2\xi|z|} \left(dz^{2}+\tilde{g}_{\mu \nu }^{(d)}(x)dx^{\mu }dx^{\nu
}\right) +\gamma _{mn}^{(10-d)}(y)dy^{m}dy^{n}, \label{Ansatz}
\end{equation}
where $\tilde{g}_{\mu \nu }^{(d)}$ stands for the worldsheet metric with $\mu
,\nu =0,...,d-1$; $\gamma _{mn}^{(10-d)}$ is the metric of $S^{10-d}$ of radius
$r_{0}$ and $\xi$ is a constant.  This Ansatz solves (\ref{R4T}) identically,
while (\ref{R5}) takes the form
\begin{equation}\label{R5+}
\epsilon_{a a_1 b_1 \cdots a_k b_k m_1 \cdots m_s}
\left[\tilde{R}^{a_1b_1}-\xi^2 \tilde{e}^{a_1}\tilde{e}^{b_1}\right] \cdots
\left[\tilde{R}^{a_k b_k1}-\xi^2 \tilde{e}^{a_k}\tilde{e}^{b_k}\right]
e^{m_1}\cdots e^{m_s}=0,
\end{equation}
where $\tilde{e}^a$ and $\tilde{R}^{ab}$ stand for the vielbein and the Riemann
curvature of the worldsheet, respectively, and $k=\left[\frac{d-1}{2}\right]$,
with $2k+s=10$.

A solution for equations (\ref{R5+}) is found if the worldsheet has constant
curvature,
\[
\tilde{R}^{ab}-\xi^{2}\,\tilde{e}^a\wedge \tilde{e}^b=0.
\]
This means that $M_{d}$ can be either locally de Sitter spacetime of radius
$\xi^{-1}$, or locally Minkowski, if $\xi=0$. The requirement that the
curvature of (\ref{Ansatz}) be a simple zero of the field equations, implies
that $k$ must be one, and therefore $d$ can only be equal to three or four. The
case $d=3$ is readily discarded as it possesses no perturbations, as in
standard three-dimensional gravity. For $d=4$, the only relevant equation for
the perturbations is the one that arises from the variation with respect to
$\tilde{e}^{i}$, the perturbations of the spacetime metric satisfy

\begin{equation} \label{d4}
\xi\delta (z)\epsilon_{ijkl}\delta
(\tilde{R}^{jk}-\xi^{2}\,\tilde{e}^j\tilde{e}^k)\tilde{e}^l=0.
\end{equation}
Since for $\xi=0$ the perturbation equations (\ref{d4}) become empty, Minkowski
spacetime must be ruled out. In sum, the existence of the propagator requires
$d=4$, and the four-dimensional cosmological constant to be strictly positive,
$\Lambda_{4}=3\xi^{2}$.

Note that Eq. (\ref{d4}) has support only on the $z=0$ plane. Perturbations
along the worldsheet, $\delta \tilde{g}_{\mu \nu}=h_{\mu \nu}(x)$ reproduce the
linearized Einstein equations in four-dimensional de Sitter spacetime. The
modes that depend on the coordinates transverse to the worldsheet fall into two
classes: those of the form $\delta \tilde{g}_{\mu\nu}=h_{\mu\nu}(x,y)$ are
massive Kaluza-Klein modes with a discrete spectrum, while $\delta
\tilde{g}_{\mu \nu }=h_{\mu \nu }(x,z)$ correspond to Randall-Sundrum-like
massive modes whose spectrum is continuous and has a mass gap. The
perturbations of the remaining metric components are zero modes, which is
related to the fact that the equations are not deterministic for the compact
space.

\section{Discussion}  

\textbf{A.} We have presented a framework in which only a four-dimensional
spacetime can sustain propagating gravitons. The mechanism is based on an
eleven--dimensional action which is a gauge theory for the M-algebra. Its is
shown that the Lagrangian can be written as a Chern-Simons form. The
possibility of dynamical dimensional reduction arises because the theory has
radically different spectra around backgrounds of different effective spacetime
dimensions. Thus, in a family of product spaces of the form $X_{d+1}
\mathbf{\times} S^{10-d}$, the only option that yields a well defined low
energy propagator for the graviton is $d=4$ and $\Lambda _{4}>0$.

It should be stressed that for all gravity theories of the type discussed here,
possessing local Poincar\'{e} invariance in dimensions $D=2n+1\geq 5$,
four-dimensional de Sitter spacetime is also uniquely selected by the same
mechanism as the background for the low--energy effective theory \cite{HTZ}.
This is in sharp contrast with the scenario in a (super)gravity theory based on
the Einstein--Hilbert Lagrangian for the gravitational sector. In the standard
cases, the kinetic term for the metric and, consequently, the propagator is
equally well behaved (no degeneracies arise) in all dimensions. In other words,
there is nothing in these theories that can produce different dynamical
behaviors for different dimensions. This may help to resolve the paradox of
compactification mentioned in the introduction: what prompts the stable
Minkowski space to curl up, loosing symmetries, and why does it compactify into
$4+(D-4)$ and not otherwise?

\textbf{B.} The eleven--dimensional action (\ref{Action}) has a free parameter
$\alpha$, which reflects the fact that the theory contains two natural limits,
corresponding to different subalgebras of (\ref{MAlgebra}). For $\alpha=0$, the
action $I_0$ in Eq. (\ref{Action}) does not depend on $b^{[5]}$ and corresponds
to a gauge theory for the supermembrane algebra, while for $\alpha=1$, the
bosonic field $b^{[2]}$ decouples, and $I_1$ is a gauge theory for the super
five-brane algebra as discussed in \cite{BTZ}. It is interesting to note that
the linear combination of both limits, $I_{\alpha }=I_{0}+\alpha (I_1-I_0)$, is
not only invariant under the intersection of both algebras, but under the
entire M-algebra. As the term $I_{1}-I_{0}$ does not couple to the vielbein and
is invariant under supersymmetry by itself, $\alpha $ is an independent
coupling constant. A similar situation occurs in nine dimensions where, in one
limit, the theory corresponds to the super five-brane algebra, while for the
other it is a gauge theory for the super-Poincar\'{e} algebra with a central
extension \cite{HOT}.

\textbf{C.} In the presence of negative cosmological constant,
eleven-dimensional AdS supergravity \cite{Tr-Z} can be written as a
Chern-Simons theory for $osp(32|1)$, which is the supersymmetric extension of
AdS$_{11}$. It is natural to ask whether there is a link between that theory in
the vanishing cosmological constant limit, and the one discussed here. Since
the M-algebra has $55$ bosonic generators more than $osp(32|1)$, these theories
cannot be
related through a In\"{o}n\"{u}-Wigner contraction for a generic value of $%
\alpha $. However, it has been recently pointed out in \cite{AIPV},
generalizing the procedure of \cite{Hatsuda-Sakaguchi}, that it is possible to
obtain the M-algebra from an expansion of $osp(32|1)$. In this light, applying
this procedure to the eleven-dimensional AdS supergravity theory, it should be
expected that the action presented here will be recovered up to some additional
terms decoupled from the vielbein, that are supersymmetric by themselves.\\
--------------------------

\textbf{Acknowledgments.-} We thank J. Edelstein, G. Kofinas, C. Mart\'{i}nez,
C. N\'{u}\~{n}ez, R. Portugu\'{e}s and C. Teitelboim for many enlightening
discussions and useful comments. This work is partially supported by grants
3020032, 1010450, 1010446, 1010449, 1020629 and 1040921 from FONDECYT. We thank
the organizers of Fourth International Winter Conference on Mathematical
Methods in Physics for the lively and stimulating conference in Rio.
Institutional support to the Centro de Estudios Cient\'{i}ficos (CECS) from
Empresas CMPC is gratefully acknowledged. CECS is a Millennium Science
Institute and is funded in part by grants from Fundaci\'{o}n Andes and the
Tinker Foundation.

\end{document}